# Who Tags What? An Analysis Framework [*]

Mahashweta Das[‡], Saravanan Thirumuruganathan[‡],
Sihem Amer-Yahia[†], Gautam Das[‡,†], Cong Yu[††]

[‡]University of Texas at Arlington; [†]Qatar Computing Research Institute; [††]Google Research

[‡]{mahashweta.das@mavs, saravanan.thirumuruganathan@mavs, gdas@cse}.uta.edu,
[†]{syahia, gdas}@qf.org.qa, [††]congyu@google.com

## ABSTRACT

The rise of Web 2.0 is signaled by sites such as Flickr, del.icio.us, and YouTube, and social tagging is essential to their success. A typical tagging action involves three components, *user*, *item* (e.g., photos in Flickr), and *tag*s (i.e., words or phrases). Analyzing how tags are assigned by certain users to certain items has important implications in helping users search for desired information. In this paper, we explore common analysis tasks and propose *a dual mining framework* for social tagging behavior mining. This framework is centered around two opposing measures, *similarity* and *diversity*, being applied to one or more tagging components, and therefore enables a wide range of analysis scenarios such as characterizing similar users tagging diverse items with similar tags, or diverse users tagging similar items with diverse tags, etc. By adopting different concrete measures for similarity and diversity in the framework, we show that a wide range of concrete analysis problems can be defined and they are NP-Complete in general. We design efficient algorithms for solving many of those problems and demonstrate, through comprehensive experiments over real data, that our algorithms significantly out-perform the exact brute-force approach without compromising analysis result quality.

## 1. INTRODUCTION

Tagging is a core activity on the social web. It reflects a wide range of content interpretations and serves many purposes, ranging from bookmarking websites in del.icio.us, organizing personal videos in YouTube, and characterizing movies in MovieLens. While one can possibly examine tags used by a single user on a single item, it is easy to see that the task becomes quickly intractable for a collection of tagging actions involving multiple users and items. In this paper, we aim to formalize the analysis of the tagging behavior of a set of users for a set of items and develop appropriate algorithms to complete that task.

A typical tagging action involves three components, *user*, *item*, and *tag*. We propose to study a variety of analysis tasks that involve applying two alternative measures, *similarity* and *diversity*, to those components and producing groups of similar or diverse items, tagged by groups of similar or diverse users with similar or diverse tags. For example, one possible analysis outcome would be: *"teenagers use diverse tags for action movies"* or *"males from New York and California use similar tags for movies directed by Cameron and Spielberg"*. In Section 2.1 and 2.2, we will describe some of these problem instances that are enabled in our framework. A general dual mining framework that encompasses many common analysis tasks is then defined in Section 2.3.

A core challenge in this dual mining framework is the design of similarity and diversity measures. For user or item components, defined by (attribute, value) pairs, several existing comparison techniques have been proposed that can leverage their structured nature or bipartite connections. Section 2.1.1 illustrates some of those techniques.

Comparing similarity and diversity of tags used by various users on different items, however, presents a new challenge. First, tags are drawn from a much larger vocabulary than user or item attributes and exhibit a long tail characteristic. Second, it is often the case that different tags are used for the same set of items and, accounting for those tags separately would not capture their co-usage. Finally, tags may have linguistic connections such as synonymy. In order to capture tag similarity and diversity, we propose to *summarize* tags first to account for their co-usage and semantic relationships. Section 2.1.2 describes some techniques from Information Retrieval and Machine Learning that can be used.

The tag component is also the most interesting among the three to be analyzed. Figure 1 shows a rendering of a tag summarization for *Woody Allen* movies in the form of a tag cloud. Similarly, Figure 2 shows a summarization of tags for the same movies from California users only. In both cases, summarization is defined as a simple frequency-based tag cloud where the size of a tag corresponds to how often it has been used on those movies. While "Woody" and "Allen" are not surprisingly common to both, the two clouds are different: all users highlight the *dramatic*, *tragic* and *disturbing* nature of those movies, and California users emphasize tags such as *classic* and *psychiatry*. Moreover, one of the direc-tor's popular movies, *Noiva Nervosa* is prominent in the tag cloud of all users, and yet is conspicuously absent in that of California users. Our goal is to define analysis tasks that can help users easily spot those interesting patterns and use that knowledge in subsequent actions.

---

[*]The work of Mahashweta Das, Saravanan Thirumuruganathan and Gautam Das is partially supported by NSF grants 0812601, 0915834, 1018865, a NHARP grant from the Texas Higher Education Coordinating Board, and grants from Microsoft Research and Nokia Research.

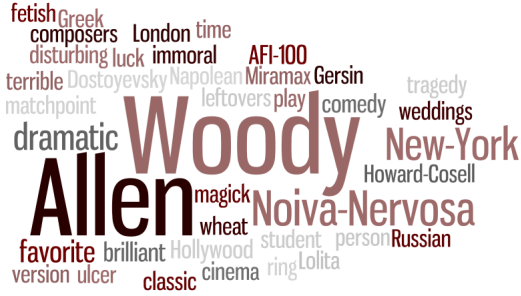

Figure 1: Tag Signature for All Users

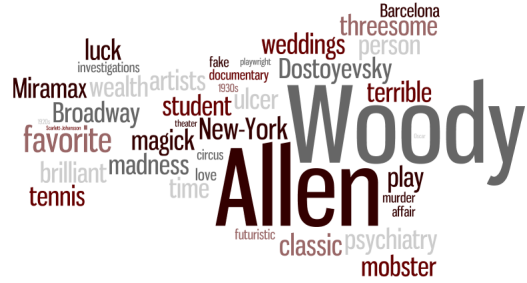

Figure 2: Tag Signature for CA Users

We emphasize that, in this study, it is *not* our goal to advocate one particular similarity or diversity measure over another. Rather, we focus on formalizing the **Tagging Behavior Dual Mining** framework and the problem definitions, and designing general algorithms that will work well for most measures.

The analysis problems formally defined in our proposed framework fall into the wider category of constrained optimization problems. We are looking for groups of tagging actions that achieve maximum similarity or diversity on one or more components while satisfying a set of constraints such as *support*. Not surprisingly, as our complexity analysis shows in Section 3, those problems are NP-Complete in general. We propose two sets of efficient algorithms for solving them. The first set incorporates Locality Sensitive Hashing (LSH) and can be used for problems maximizing tagging action component similarity. While traditional LSH is frequently used for performing nearest neighbor search in high-dimensional spaces, our algorithm finds the bucket containing the result set of our tagging behavior analysis. The second set of algorithms borrows ideas from techniques employed in Computational Geometry to handle the Facility Dispersion Problem (FDP) and is effective for problems maximizing diversity. Both sets of algorithms possess compelling theoretical characteristics for problem instances optimizing the dual mining goal without any constraints. For both sets, we also propose advanced techniques that return better quality results in comparable running time.

In summary, we make the following main contributions:

- We formalize the task of analyzing the tagging behavior of a set of users for a set of items and propose a novel general constrained optimization framework for tagging behavior mining.
- We show that the tagging analysis problems are NP-Complete and propose efficient algorithms for solving the problems.
- We develop locality sensitive hashing based algorithms for solving problems maximizing tagging action component similarity. We also design computational geometry based algorithms for problem instances maximizing diversity. We provide theoretical guarantees for both sets of algorithms for handling problems optimizing the dual mining goal without any constraints.

- We perform detailed experiments on real data to show that our proposed algorithms generate equally good results as exact brute-force in much less execution time.

## 2. THE TAGDM FRAMEWORK

We model the data on a social tagging site as a triple $\langle U, I, T \rangle$, representing the set of users, the set of items and the tag vocabulary, respectively. Each tagging action can be considered as a triple itself, represented as $\langle u, i, \texttt{T} \rangle$, where $u \in U$, $i \in I$, $\texttt{T} \subset T$, respectively. A group of tagging actions is denoted as $g = \{\langle u_1, i_1, \texttt{T}_1 \rangle, \langle u_2, i_2, \texttt{T}_2 \rangle, \ldots, \}$. We define a user schema, $S_U = \langle a_1, a_2, \ldots \rangle$, to represent each user as a set of attribute values conforming to the user schema: $u = \langle u.a_1, u.a_2, \ldots \rangle$, where each $u.a_x$ is a value for the attribute $a_x \in S_U$. For example, let $S_U = \langle \texttt{age}, \texttt{gender}, \texttt{state}, \texttt{city} \rangle$, a user can be represented as $\langle \texttt{18}, \texttt{student}, \texttt{new york}, \texttt{nyc} \rangle$. Similarly, we define an item schema, $S_I = \langle a_1, a_2, \ldots \rangle$, to represent each item as a set of attribute values, $i = \langle i.a_1, i.a_2, \ldots \rangle$, where each $i.a_y$ is a value for the attribute $a_y \in S_I$.

Each tagging action therefore can be represented as an expanded tuple that concatenates the user attributes, the item attributes and the tags: $r = \langle r_u.a_1, r_u.a_2, \ldots, r_i.a_1, r_i.a_2, \ldots, \texttt{T} \rangle$. $G$ denotes the set of all such tagging action tuples. Many social sites have hundreds of millions of such tuples. Most, if not all, mining tasks involve analyzing sets of such tuples collectively. While there are a number of different ways tagging action tuples can be grouped, we adopt the view proposed and experimentally verified in [6], where groups of users (or items) that are *structurally describable* (i.e., sharing common attribute value pairs) are meaningful to end-users. Such groups correspond to conjunctive predicates on user or item attributes. An example of a *user describable tagging action group* is {gender=male, state=new york}, and of an *item describable group* is {genre=comedy, director=woody allen}. Next we define an essential characteristic of a set of tagging action groups.

DEFINITION 1. **Group Support**. *Given the input set of tagging action tuples $G$, the support of a set of tagging action groups $\mathcal{G} = \{g_1, g_2, \ldots\}$ over $G$, is defined as $Support_G^{\mathcal{G}} = |\{r \in G \mid \exists g_x \in \mathcal{G}, r \in g_x\}|$.* Intuitively, group support measures the number of input tagging action tuples that belongs to at least one of the groups in $\mathcal{G}$.

### 2.1 Concrete TagDM Problems

A large number of concrete *TagDM* problem instances can be defined, with their variations coming from two main aspects. The first category of variations depends on which

1568

measure, similarity or diversity, the user is interested in applying to which tagging components (i.e, users, items, or tags). For example, a user can be interested in identifying *similar* tags produced by *similar* user groups on *diverse* item groups, or *similar* tags produced by *diverse* user groups on *similar* item groups. Since there are three components, each of which can adopt one of two measures, this variation alone can lead to $2^3 = 8$ different problem instances.

The second category of variations depends on which components the user is adding to the optimization goal and which components the user is adding to the constraints. For example, a user can be interested in finding tagging action groups that maximize a *tag diversity* measure and satisfy *user and item similarity* constraints, or groups that maximize a combination of *tag diversity and user diversity* measures and satisfy an *item similarity* constraint. Since each component can be part of the optimization goal, or part of the constraint, or neither, this variation can lead to $3^3 - 1 = 26$ different problem instances.

Combining both categories of variations, there is a total of **112** concrete problem instances that our framework captures! Table 1 illustrates six of the problem instantiations that we have investigated in detail. In particular, we focus on problems with all three components with constraints on user and item and optimization on the tag component, since those are the most novel and intuitive mining problems.

| ID | User | Item | Tag | $C$ | $O$ |
|---|---|---|---|---|---|
| 1 | similarity | similarity | similarity | U,I | T |
| 2 | similarity | diversity | similarity | U,I | T |
| 3 | diversity | similarity | similarity | U,I | T |
| 4 | diversity | similarity | diversity | U,I | T |
| 5 | similarity | diversity | diversity | U,I | T |
| 6 | similarity | similarity | diversity | U,I | T |

**Table 1: Concrete *TagDM* Problem Instantiations. Column $C$ lists the constraint dimensions and column $O$ lists the optimization dimensions.**

Before we formalize the mining problems, we introduce the core concept of *Dual Mining Function*.

DEFINITION 2. **Dual Mining Function**. *A Dual Mining Function, $F : \mathcal{G} \times b \times m \to$ float, takes as inputs: $\mathcal{G}$, a set of tagging action groups; $b \in \{\texttt{users}, \texttt{items}, \texttt{tags}\}$, a tagging behavior dimension; $m \in \{\texttt{similarity}, \texttt{diversity}\}$, a dual mining criterion; and produces a float score, $s$, that quantifies the mining criterion over the particular dimension for the set of tagging action groups.*

Definition 2 defines a general dual mining function that computes a score using arbitrary evaluations over the tagging action groups. In practice, there is a subset of dual mining functions that are more restricted and yet powerful enough for solving many real mining scenarios:

DEFINITION 3. **Pair-Wise Aggregation Dual Mining Function**. *A Pair-Wise Aggregation (PA) Dual Mining Function, $F_{pa} : \mathcal{G} \times b \times m \to$ float, is a dual mining function with two component function $F_p : g_i \times g_j \times b \times m \to$ float and $F_a : \{s_1, s_2, \ldots\} \to$ float, where $(g_i, g_j)$ is a pair of distinct tagging action groups and each $s_i$ is an intermediate score produced by $F_p$, such that: $F_{pa}(\mathcal{G}, b, m) = F_a(\{F_p(g_i, g_j, b, m)\}, \forall g_i, g_j \in \mathcal{G}, i \neq j$.*

Pair-wise dual mining functions simplify the general dual mining functions by enabling the overall mining score to be computed via aggregating the scores computed over pairs of the tagging action groups, which is often much easier to define and compute. We now present a few examples of the pair-wise dual mining function. The key to a pair-wise dual mining function is the pair-wise comparison function, $F_p(g_1, g_2, b, m)$, where $g_1$ and $g_2$ are distinct tagging action groups, and $b \in \{\texttt{users}, \texttt{items}, \texttt{tags}\}$, is a tagging behavior dimension, and $m \in \{\texttt{similarity}, \texttt{diversity}\}$, is a dual mining criterion.

### 2.1.1 User & Item Dimensions Dual Mining

Given a user describable tagging action group[1], its user dimension is effectively its user group description, i.e., a set of (attribute, value) pairs that describes the group. Therefore, given two user groups, $g_1$ and $g_2$, their similarity or diversity can be captured mainly in two ways: 1) structural distance between the user group descriptions and 2) set distance based on the items they have rated.

Let $A$ be the set of user attributes shared between two user describable tagging action groups $g_1$ and $g_2$, an example of the pair-wise comparison function leveraging **structural distance** is the following:

$F_p(g_1, g_2, \texttt{users}, \texttt{similarity}) = \sum_{a \in A} sim(v_1, v_2)$

where $a.v_1$ and $a.v_2$ belong to the set of user attribute value pairs and $sim$ can be a string similarity function that simply computes the edit distance between two values or a more sophisticated similarity function that takes domain knowledge into consideration. For example, a domain-aware similarity function can determine "New York City" to be more similar to "Boston" than to "Dallas". $F_p(g_1, g_2, \texttt{users}, \texttt{diversity})$ can be similarly defined using the inverse function.

Let $g_1.I$ and $g_2.I$ be the sets of items tagged by tuples in $g_1$ and $g_2$, respectively, an example of the pair-wise comparison function leveraging **set distance** is the following:

$F'_p(g_1, g_2, \texttt{users}, \texttt{similarity}) = \frac{|\{r|r \in g_1.I \wedge r \in g_2.I\}|}{|\{r|r \in g_1.I \vee r \in g_2.I\}|}$

which simply computes the percentages of items tagged by both groups (akin to Jaccard distance.) If numerical ratings are available for each tagging tuple, a more sophisticated set distance similarity function can further impose an additional constraint that an item is common to both groups if its average ratings in both are close. $F'_p(g_1, g_2, \texttt{users}, \texttt{diversity})$ can be similarly defined using the inverse function.

### 2.1.2 Tag Dimension Dual Mining

The tag dimension is fundamentally different from the user and item dimensions. First, there is no fixed set of attributes associated with the tag dimension, therefore the structural distance does not apply. Second, tags are chosen freely by users using diverse vocabularies. As a result, a single tagging action group can contain a large number of tags. Both characteristics make comparing two sets of tags very difficult.

We propose a two-step approach for handling the tag dimension. First, we propose an initial step to summarize the set of all tags of a tagging action group into a smaller representative set of tags, called *group tag signature*. Second, we apply comparison functions to compute distance between signatures. Once again, we are not advocating any particular way of producing signatures and/or comparing them. Rather, we simply argue for the need for tag signatures and their comparisons.

---

[1]Since the user and item dimensions share the same characteristics in the dual mining framework, we present here only the user dimension for simplicity.

**Group tag signature generation**: Given a group of tagging actions $g = \{\langle u_1, i_1, \texttt{T}_1\rangle, \langle u_2, i_2, \texttt{T}_2\rangle, \ldots\}$, we aim to summarize the tags in $\texttt{T}_1 \cup \texttt{T}_2 \cup \ldots$ into a tag signature $T_{rep}(g)$. The general form of $T_{rep}(g)$ is $\{(tc_1, w_1), (tc_2, w_2), \ldots\}$ where $tc_i$ is topic category (can be a tag itself) and $w_i$ is weight, i.e., relevance of $g$ for $c_i$.

One can define several methods to compute tag signatures. For example, when tags are hand-picked by editors and hence the number of unique tags is small, a simple definition can be $T_{rep}(g) = \{(t, \texttt{freq}(t)) \mid t \in \texttt{T}_1 \cup \texttt{T}_2 \cup \ldots\}$, where $\texttt{freq}(t)$ computes how many times $t$ is used in $g$.

Most collaborative tagging sites encourage users to create their own tags, thereby creating a long tail of tags. This raises challenges such as sparsity and the choice of different tags to express similar meanings. Techniques from Information Retrieval and Machine Learning such as tf*idf and Latent Dirichlet Allocation (LDA) can be used for tag summarization. LDA aggregates tags into topics based on their co-occurrence and reason at the level of topics, and handles long tail issues [2]. Also, a Web service such as Open Calais[2] can be used to match a set of tags to a set of predefined categories through sophisticated language analysis and information extraction.

**Comparing group tag signatures**: When tagging action groups are represented as tag signatures over the same set of topics, we can leverage many existing vector comparison functions to compute the distance between any two group tag signature vectors pair-wisely. An example is simply **cosine similarity** as follows:
$F''_P(g_1, g_2, \texttt{tags}, \texttt{similarity}) = \cos(\theta(T_{rep}(g_1), T_{rep}(g_2)))$, where $\theta$ is the angle between the two vectors. $F''_P(g_1, g_2, \texttt{tags}, \texttt{diversity})$ can be defined similarly

The comparison can also be enhanced by using an ontology such as WordNet to compare entries of similar topics.

## 2.2 Common Problem Instances

We are now ready to define two of the concrete dual mining problems listed in Table 1. The first one aims to find similar user sub-populations who agree most on their tagging behavior for a diverse set of items. The second one aims to find diverse user sub-populations who disagree most on their tagging behavior for a similar set of items.

PROBLEM 1. *Identify a set of tagging action groups, $G^{opt} = \{g_1, g_2, \ldots\}$, such that:*
- $\forall g_x \in G^{opt}$, $g_x$ *is user- and/or item-describable;*
- $1 \leq |G^{opt}| \leq k$;
- $Support_G^{G^{opt}} \geq p$;
- $F_1(G^{opt}, \texttt{users}, \texttt{similarity}) \geq q$;
- $F_2(G^{opt}, \texttt{items}, \texttt{diversity}) \geq r$;
- $F_3(G^{opt}, \texttt{tags}, \texttt{similarity})$ *is maximized.*

where $F_1$ and $F_2$ are structural similarity based dual mining functions as defined in Definition described in Section 2.1.1, and $F_3$ is the LDA based tag dual mining function as described in Section 2.1.2.

For $k = 2$, $p = 100$, $q = 0.5$, and $r = 0.5$, solving the problem on the full set of tagging action tuples in MovieLens [5] can give us the following $G^{opt}$:
$g_1 = \{\langle \texttt{gender}, \texttt{male}\rangle, \langle \texttt{age}, \texttt{young}\rangle, \langle \texttt{actor}, \texttt{j.aniston}\rangle,$
$\quad (comedy, drama, friendship)\}$
$g_2 = \{\langle \texttt{gender}, \texttt{male}\rangle, \langle \texttt{age}, \texttt{young}\rangle, \langle \texttt{actor}, \texttt{j.timberlake}\rangle,$
$\quad (drama, friendship)\}$

[2]*https://www.opencalais.com/*

which illustrates the interesting pattern that male young users assign similar tags, *drama* and *friendship*, to movies with "Jennifer Aniston" and "Justin Timberlake," the former for her involvement in the popular TV show "Friends" and the latter for his movie "The Social Network."

A closely related problem to Problem 1 is to inverse the similarity and diversity constraints for the user and item components, i.e., finding diverse user sub-populations who agree most on their tagging behavior for a similar set of items (Problem 3 in Table 1). Both problems focus on optimizing the tag similarity and therefore can be solved using similar techniques. Next, we define a problem that aims to identify groups that disagree on their tagging behavior.

PROBLEM 4. *Identify a set of tagging action groups, $G^{opt} = \{g_1, g_2, \ldots\}$, such that:*
- $\forall g_x \in G^{opt}$, $g_x$ *is user- and/or item-describable;*
- $1 \leq |G^{opt}| \leq k$;
- $Support_G^{G^{opt}} \geq p$;
- $F_1(G^{opt}, \texttt{users}, \texttt{diversity}) \geq q$;
- $F_2(G^{opt}, \texttt{items}, \texttt{similarity}) \geq r$;
- $F_3(G^{opt}, \texttt{tags}, \texttt{diversity})$ *is maximized.*

where $F_1$, $F_2$, and $F_3$ are similarly defined as in Problem 1.

For $k = 2$, $p = 100$, $q = 0.5$, and $r = 0.5$, solving the problem on the full set of tagging action tuples in MovieLens can give us the following $G^{opt}$:
$g_1 = \{\langle \texttt{gender}, \texttt{male}\rangle, \langle \texttt{age}, \texttt{teen}\rangle, \langle \texttt{genre}, \texttt{action}\rangle,$
$\quad (gun, special\ effects)\}$
$g_2 = \{\langle \texttt{gender}, \texttt{female}\rangle, \langle \texttt{age}, \texttt{teen}\rangle, \langle \texttt{genre}, \texttt{action}\rangle,$
$\quad (violence, gory)\}$

which illustrates teenaged male users and female users have entirely different perspectives on action movies. This gives a user a new insight that there is something about action movies that is causing the different reactions among two different groups of users.

## 2.3 Generalizing the TagDM Framework

We take the novel approach of proposing a general constrained optimization framework for tagging behavior mining, upon which various analysis tasks can be instantiated and optimized.

DEFINITION 4. **Tagging Behavior Dual Mining (TagDM) Problem**. *Given a triple $\langle G, C, O \rangle$ in the TagDM framework where $G$ is the input set of tagging actions and $C$, $O$ are the sets of constraints and optimization criteria respectively, the Tagging Behavior Dual Mining problem is to identify a set of tagging action groups, $G^{opt} = \{g_1, g_2, \ldots\}$ for $b \in \{\texttt{users}, \texttt{items}, \texttt{tags}\}$ and $m \in \{\texttt{similarity}, \texttt{diversity}\}$, such that:*
- $\forall g_x \in G^{opt}$, $g_x$ *is user- and/or item-describable;*
- $k_{lo} \leq |G^{opt}| \leq k_{hi}$;
- $Support_G^{G^{opt}} \geq p$;
- $\forall c_i \in C, c_i.F(G^{opt}, b, m) \geq threshold$;
- $\Sigma_{o_j \in O}\, o_j.F(G^{opt}, b, m)$ *is maximized.*

Intuitively, TagDM aims to identify a set of user- and/or item-describable sub-groups from input tagging actions, such that the dual mining constraints are satisfied and a dual mining goal is optimized. We now clearly see how this framework generalizes the common problem instances given in Section 2.2.

## 3. COMPLEXITY ANALYSIS

In this section we provide the proof that the Tagging Behavior Dual Mining problem is NP-Complete. The decision version of the TagDM problem is defined as follows:

Given a triple $\langle G, C, O \rangle$, is there a set of tagging action groups $G^{opt} = \{g_1, g_2, \ldots\}$ such that $\sum_{o_j \in O}(o_j.Wt \times o_j.F(G_{opt}, o_j.D, o_j.M)) \geq \alpha$ subject to:

- $\forall g_x \in G^{opt}, g_x$ is user- and/or item-describable.
- $k_{lo} \leq |G^{opt}| \leq k_{hi}$
- $Support_G^{G^{opt}} \geq p$
- $\forall c_i \in C, c_i.F(G^{opt}, c_i.D, c_i.M) \geq c_i.Th$

THEOREM 1. *The decision version of the TagDM problem is NP-Complete.*

PROOF. The membership of decision version of TagDM problem in NP is obvious. To verify NP-Completeness, we reduce Complete Bipartite Subgraph problem (CBS) to our problem and argue that a solution to CBS exists, *if and only if*, a solution our instance of TagDM exists. First, we show that the problem CBS is NP-Complete.

LEMMA 1. *Complete bipartite subgraph problem (CBS) is NP-Complete.*

PROOF. The decision version of CBS is defined as follows:

Given a bipartite graph $G' = (V_1, V_2, E)$ and two positive integers $n_1 \leq |V_1|, n_2 \leq |V_2|$, are there two disjoint subsets $V_1' \subseteq V_1, V_2' \subseteq V_2$ such that $|V_1'| = n_1, |V_2'| = n_2$ and $u \in V_1', v \in V_2'$ implies that $\{u, v\} \in E$.

The membership of CBS in NP is obvious. We verify the NP-Completeness of the problem by reducing it to Balanced Complete Bipartite Subgraph (BCBS) problem which is defined as : Given a bipartite graph $G'' = (V_1'', V_2'', E')$ and a positive integer $n'$, find two disjoint subsets $V_1''' \subseteq V_1'', V_2''' \subseteq V_2''$ such that $|V_1'''| = |V_2'''| = n'$ and $u \in V_1''', v \in V_2'''$ implies that $\{u, v\} \in E'$. This problem was proved to be NP-Complete by reduction from Clique in [14]. We can reduce BCBS to CBS by passing the input graph $G''(V_1'', V_2'', E')$ of BCBS to CBS and setting $n_1$ and $n_2$ to $n'$. If a solution exists for the CBS instances, then the disjoint subsets $V_1''', V_2'''$ form a balanced complete bipartite subgraph in $G''$. □

We have already established that TagDM problem is in NP. To verify its NP-Completeness, we reduce CBS to the decision version of our problem. Given an instance of the problem CBS with $G' = (V_1, V_2, E)$ and positive integers $n_1, n_2$, we construct an instance of TagDM problem such that there exists a complete bipartite subgraph induced by disjoint vertex subsets $V_1' \subseteq V_1, V_2' \subseteq V_2$ and $|V_1'| = n_1, |V_2'| = n_2$, *if and only if*, a solution to our instance of TagDM exists.

First, we create an user schema $S_U = \langle a_1, a_2, \ldots, a_{|V_2|} \rangle$ such that for each vertex $v_j \in V_2$, there exists a corresponding user attribute $a_j \in S_U$. Next, we define a set of users $U = \{u_1, u_2, \ldots, u_{|V_1|}\}$. Again, for each vertex $v_i \in V_1$ there exists a corresponding user $u_i \in U$.

For all pairs of vertices $(v_i, v_j), v_i \in V_1, v_j \in V_2$, we set $u_i.a_j$ to 1 if $\{v_i, v_j\} \in E$; else, we set it to a unique value such that $u_{x1}.a_{y1} \neq u_{x2}.a_{y2}$ unless $x_1 = x_2, y_1 = y_2$. Intuitively, we set the j-th attribute of i-th user to 1 if an edge exists between vertex pairs $(v_i, v_j)$; else, we set it to a unique value that is not shared with any attribute of any user. One possible way to assign the unique attribute values is to pick a previously unassigned value from the set $[2, |V_1| \times |V_2| + 1]$. Since the number of possible edges is at most $|V_1| \times |V_2|$, this set suffices to generate unique attribute values.

We construct an instance of the TagDM problem where $I = \{i\}$ and $T = \{t\}$. This results in a set of tagging actions, $G = \{\langle u_1, i, t\rangle, \ldots, \langle u_{|V_1|}, i, t \rangle\}$ where only the user dimension plays a non-trivial role in determining the problem solution. Given a pair of users, the pairwise similarity function $F_1$ on user dimension measures their structural similarity by counting the number of attribute values that are shared between them. Intuitively, the problem collapses to that of finding a subset of users who share a subset of attributes.

We then define our TagDM problem instance as : For a given a triple $\langle G, C, O \rangle$, identify a set $G^{opt}$ of tagging action groups such that $F_3(G^{opt}, tags, m) \geq 0$ subject to:

- $1 \leq |G^{opt}| \leq n_1$
- $Support_G^{G^{opt}} \geq n_1$
- $F_1(G^{opt}, \texttt{users}, \texttt{similarity}) \geq n_2 \times \binom{n_1}{2}$

If there exists a solution to this TagDM problem instance, then there are $n_1$ users who have identical values for at least $n_2$ of their attributes. If two users $u_x$ and $u_y$ have same values for a set of attributes $A$, then for all attributes $a \in A$, $u_x.a = u_y.a = 1$. In other words, whenever the attributes of two users overlap, the shared attributes can only take a value of 1. Any other symbol that was assigned is unique and cannot overlap by construction. If there exists a subset of attributes $A' \subseteq S_U$ and a subset of users $U' \subseteq U$, then the corresponding vertices in $V_1$ and $V_2$ form a complete bipartite subgraph solving the input instance of BCS. Thus TagDM problem is NP-Complete. □

### 3.1 Exact Algorithm

A brute-force exhaustive approach (henceforth, referred to as **Exact**) to solve the TagDM problem requires us to enumerate all possible combinations of tagging action groups in order to return the optimal set of groups maximizing the mining criterion and satisfying the constraints. The number of possible candidate sets is exponential in the number of groups. Evaluating the constraints on each of the candidate sets and selecting the optimal result can thus be prohibitively expensive. Each tagging action group is associated with a group tag signature vector (the size of which is determined by the cardinality of the global set of topics), which may introduce additional challenges in the form of higher dimensionality. Therefore, we develop practical and efficient algorithms.

We develop two sets of algorithms. The first set comprises of locality sensitive hashing based algorithms for handling TagDM problem instances maximizing similarity of tagging action components. The algorithms are efficient in practice, but cannot handle TagDM problem instances maximizing diversity. The second set is based on techniques employed in computational geometry for the facility dispersion problem and is our solution for diversity mining problem instances.

## 4. LSH BASED ALGORITHMS

The first of our algorithmic solutions is based on locality sensitive hashing (**LSH**) which is a popular technique to solve nearest neighbor search problems in higher dimensions [13]. The basic idea is to hash similar input items into the same bucket (i.e., uniquely definable hash signature) with high probability. It performs probabilistic dimension reduction of high dimensional data by projecting

1571

input items in higher dimension to a lower dimension such that items that were in close proximity in the higher dimension get mapped to the same item in the lower dimensional space with high probability. LSH guarantees a lower bound on the probability that two similar input items fall into the same bucket in the projected space and also the upper bound on the probability that two dissimilar vectors fall into the same bucket. For any pair of points in a high-dimensional space, $P_1$ is the probability of two close points falling into the same bucket and $P_2$ is the probability of two far-apart points falling into the same bucket; we want $P_2 < P_1$. If input items are projected from higher dimension $d$ to a lower dimension $d'$, the probabilities can be *bounded* by:

$$P(similar\ items\ colliding) \geq (1-P_1)^{d'} \quad (1)$$
$$P(dissimilar\ items\ colliding) \leq P_2^{d'}$$

This provides an approach to select a set of tagging action groups that are similar in their tagging behavior. In our problem, we need to compare the input set $G$ of $n$ tagging action groups (i.e., $n$ $d$-dimensional tag signature vectors, where $d$ is the cardinality of the global set of tag topic categories mentioned in Section 2.1.2) using a pairwise comparison function $F_p''(g_1, g_2, \texttt{tags}, \texttt{similarity})$ that operates on group tag signature vectors in order to optimize tag similarity. The result set of tagging action groups $G^{opt}$ maximizing tag similarity can be retrieved by finding the $k$ closest vectors with minimum average pairwise distance between them.

Note that, our LSH based algorithms works for Problems 1, 2 and 3 in Table 1 maximizing tag similarity. We first introduce an algorithm that returns the set of tagging action groups $G^{opt}$, $1 \leq |G^{opt}| \leq k$ having maximum similarity in tagging behavior (Column $O$ in Table 1) and then discuss additional techniques to include the multiple hard constraints into the solution (Column $C$ in Table 1).

### 4.1 Maximizing Similarity based on LSH

Our LSH based algorithm **SM-LSH** deals with TagDM problem instances optimizing tag **Si**Milarity. In traditional LSH, the buckets obtained after hashing input items are used to find the nearest neighbors for new items. In our solution, we instead rank the buckets based on the scoring function. One of the key requirement for good performance of LSH is the careful selection of the family of hashing functions. In SM-LSH, we use the LSH scheme proposed by Charikar [4] which employs a family of hashing functions based on cosine similarity. As discussed in Section 2.1.2, the cosine similarity between two tagging action group tag signature vectors is defined as the cosine of the angle between them and can be defined as:

$$cos(\theta(T_{rep}(g_x), T_{rep}(g_y))) = \frac{|T_{rep}(g_x).T_{rep}(g_y)|}{\sqrt{|T_{rep}(g_x)|.|T_{rep}(g_x)|}}$$

The algorithm computes a succinct hash signature of the input set of $n$ tagging action groups by computing $d'$ independent dot products of each $d$-dimensional group tag signature vector $T_{rep}(g_x)$, where $g_x \subseteq G$ with a random unit vector $\vec{r}$ and retaining the sign of the $d'$ resulting products. This maps a higher $d$-dimensional vector to a lower $d'$-dimensional vector ($d' << d$). Each entry of $\vec{r}$ is drawn from a 1-dimensional Normal distribution N(0,1) with zero mean and unit variance. Alternatively, we can generate a spherically symmetric random vector $\vec{r}$ of unit length from the $d$-dimensional space. The LSH function for cosine similarity for our problem is given by the following Theorem 2 [4]:

THEOREM 2. *Given a collection of n d-dimensional vectors where each vector $T_{rep}(g_x)$ corresponds to a $g_x \subseteq G$, and a random unit vector $\vec{r}$ drawn from a 1-dimensional Normal distribution N(0,1), define the hash function $h_r$ as:*
$$h_r(T_{rep}(g_x)) = \begin{cases} 1 & if\ \vec{r}.T_{rep}(g_x) \geq 0 \\ 0 & if\ \vec{r}.T_{rep}(g_x) < 0 \end{cases}$$
*Then for two arbitrary vectors $T_{rep}(g_x)$ and $T_{rep}(g_y)$, the probability that they will fall in the same bucket is:*
$$P[h_r(T_{rep}(g_x)) = h_r(T_{rep}(g_y))] = 1 - \frac{\theta(T_{rep}(g_x), T_{rep}(g_y))}{\pi}$$
*where $\theta(T_{rep}(g_x), T_{rep}(g_y))$ is angle between two vectors.*

The proof of the above Theorem 2 establishing that the probability of a random hyperplane (defined by $\vec{r}$ to hash input vectors) separating two vectors is directly proportional to the angle between the two vectors follows from Goemans et. al's theorem [9]. Any pairwise dual mining function for comparing tag signatures must satisfy such properties. We represent the $d'$-dimensional-bit LSH function as:
$$g(T_{rep}(g_x)) = [h_{r_1}(T_{rep}(g_x)), \ldots, h_{r_{d'}}(T_{rep}(g_x))]^T$$

For $d'$ LSH functions and from (1), the probability of similar tag signature vectors falling into the same bucket for all $d'$ hash functions is given by:

$$P(similar\ tag\ vectors\ colliding) \geq \left(\frac{\theta(T_{rep}(g_x), T_{rep}(g_y))}{\pi}\right)^{d'}$$

Now, each input vector is entered into $l$ hash tables indexed by independently constructed hash functions $g_1(T_{rep}(g_x)), \ldots, g_l(T_{rep}(g_x))$. Using this LSH scheme, we hash the group tag signature vectors to $l$ different $d'$-dimensional hash signatures(or, buckets). The total number of possible hash signatures in each of the $l$ lower dimensional space is $2^{d'}$. However, the maximum bound on the number of buckets in each of the lower dimensional space is $n$.

While LSH is generally used to find the nearest neighbors for new items, we take the novel approach of finding the right bucket to output as result of our problem based on checking for the number of tagging action groups in result set and ranking by scoring function. For each of the $l$ hash tables, we first check for satisfiability of $1 \leq |G^{opt}s| \leq k$ in each bucket and then rank the buckets based on the scoring function in order to determine the result set of tagging action groups $G^{opt}$ with maximum similarity.

THEOREM 3. *Given a collection of n d-dimensional tag signature vectors where each pair of vectors $T_{rep}(g_x)$ and $T_{rep}(g_y)$ corresponds to a $g_x, g_y \subseteq G$, the probability of finding result set $G^{opt}$ of k most similar vectors by SM-LSH is bounded by:*
$$P(G^{opt}) \geq 1 - \sum_{x,y \in [1,k]} [1 - \left(\frac{\theta(T_{rep}(g_x), T_{rep}(g_y))}{\pi}\right)^{d'}]$$

PROOF. The probability of finding the set of tagging action groups $G^{opt}, 1 \leq |G^{opt}| \leq k$ having maximum similarity in tagging behavior, $P(G^{opt})$:
= 1 - P(one of $^kC_2$ vector pair belongs to different buckets)
$\geq$ 1 - $\sum_{x,y \in [1,k]}$ P($T_{rep}(g_x), T_{rep}(g_y)$ in different buckets)
$\geq$ 1 - $\sum_{x,y \in [1,k]}$ [ 1 - P($T_{rep}(g_x), T_{rep}(g_y)$ in same buckets) ]
$\geq$ 1 - $\sum_{x,y \in [1,k]}$ [ 1 - $\left(\frac{\theta(T_{rep}(g_x), T_{rep}(g_y))}{\pi}\right)^{d'}$ ]  □

The above theorem establishes the theoretical probabilistic bound of finding the optimal result set. This is a Monte Carlo randomized algorithm whose probability of success can be boosted by either increasing the number of hash functions $d'$ or the number of trials of the algorithm. We also validate the efficiency of our technique in a practical setting in Section 6.

Algorithm 1 is the pseudo code of our SM-LSH algorithm. This algorithm may return null result if post-processing of all $l$ hash tables yields no bucket satisfying $1 \leq |G^{opt}| \leq k$. This could be either because there are no set of tagging action groups that satisfy the $\leq k$ requirement of our problem instance or because the input parameters to LSH caused a partitioning of data that seperated the candidate set of groups across different buckets. This motivates us to tune SM-LSH by *iterative relaxation* that varies the input parameter $d'$ in each iteration. Decreasing the parameter $d'$ increases the expected number of tagging action groups hashing into a bucket, thereby increasing the chances of our algorithm finding the result set. We perform a binary search between 1 and $d'$ to identify the correct number of hash functions to employ.

**Complexity Analysis:** The pre-processing or locality sensitive hashing time is bounded by $O(nld' \log n)$ since the binary search relaxation iteration runs for $\log n$ times in the worst case and hashing time is $O(nld')$. In the second phase, we post-process the buckets for ranking by scoring function which is a $O(n \log n)$ operation. The space complexity of the algorithm is $O(nl)$ since there are $l$ hash tables and each table has at most $n$ buckets.

SM-LSH is a fast algorithm with interesting probabilistic guarantees and is advantageous, especially for high-dimensional input vectors. However, the hard constraints along user and item dimensions are not leveraged in the optimization solution so far. Next, we discuss LSH based approaches for accommodating the multiple hard constraints into the solution.

### 4.2 Dealing with Constraints: Filtering

A straightforward method of refining the result set of SM-LSH for satisfiability of all the hard constraints in TagDM problem instances is post-processing or **Fi**ltering. We refer to this algorithm as **SM-LSH-Fi**. For each of the $l$ hash tables, we first check for satisfiability of the hard constraints in each bucket and then rank the buckets (satisfying hard constraints) according to the scoring function in order to determine the result set of tagging action groups $G^{app}$ (We represent $G^{opt}$ as $G^{app}$ since LSH based technique now perform approximate nearest neighbor search) with maximum similarity. Such post-processing of buckets for satisfiability of hard constraints may also return null results, if post-processing of hash tables yields no bucket satisfying all the hard constraints. Therefore, we propose a smarter method that folds the hard constraints concerning similarity as part of vectors in high-dimensional space, thereby increasing the chances of similar groups hashing into the same bucket.

### 4.3 Dealing with Constraints: Folding

Problems 2 and 3 in Table 1 has two out of the three tagging action components to be mined for similarity. In order to explore the main idea of LSH, we **Fo**ld the hard constraints maximizing similarity as soft constraints into our SM-LSH algorithm in order to hash similar input tagging action groups (similar with respect to group tag signature vector and user and/or item attributes) into the same bucket with high probability. We refer to this algorithm as **SM-LSH-Fo**. We fold the user or item similarity hard constraints in Problems 2 and 3 respectively into the optimization goal and apply our algorithm, so that tagging

**Algorithm 1** SM-LSH ($G$, $O$, $k$, $d'$, $l$): $G^{opt}$

*//Main Algorithm*
1: $min = 1$
2: $max = d'$
3: $T^U_{rep} \leftarrow \{\}; T^I_{rep} \leftarrow \{\}$
4: **if** $C_1.m = \texttt{similarity}$ **then**
5:     $T^U_{rep} \leftarrow$ Unarize user vector
6: **end if**
7: **if** $C_2.m = \texttt{similarity}$ **then**
8:     $T^I_{rep} \leftarrow$ Unarize item vector
9: **end if**
10: **for** $x = 1$ to $n$ **do**
11:     $T_{rep}(g_x) \leftarrow T^U_{rep}(g_x) + T^I_{rep}(g_x) + T_{rep}(g_x)$
12: **end for**
13: **repeat**
14:     $Buckets \leftarrow$ LSH($G$, $d'$, $l$)
15:     $G^{opt} \leftarrow$ MAX(Rank($Buckets$, $k$))
16:     **if** $G^{opt} = null$ **then**
17:         $max = d' - 1$
18:     **else**
19:         $min = d' + 1$
20:     **end if**
21:     $d' = (min + max)/2$
22: **until** ($min > max$) **or** ($G^{opt} \neq null$)
23: **return** $G^{opt}$

*//LSH(G, d', l): Buckets*
1: **for** $z = 1$ to $l$ **do**
2:     **for** $x = 1$ to $n$ **do**
3:         **for** $y = 1$ to $d'$ **do**
4:             Randomly choose $\vec{r}$ from $d$-dimensional Normal distribution N(0, 1)
5:             **if** $\vec{r}.T_{rep}(g_x) \geq 0$ **then**
6:                 $h_{r_y}(T_{rep}(g_x)) \leftarrow 1$
7:             **else**
8:                 $h_{r_y}(T_{rep}(g_x)) \leftarrow 0$
9:             **end if**
10:            $g_z(T_{rep}(g_x)) = [h_{r_1}(T_{rep}(g_x)), .., h_{r_{d'}}(T_{rep}(g_x))]^T$
11:        **end for**
12:    **end for**
13: **end for**
14: $Buckets \leftarrow g_1(T_{rep}(g_x)) \cup \cdots \cup g_l(T_{rep}(g_x))$
15: **return** $Buckets$

action groups with similar user attributes or similar item attributes, and similar group tag signature vectors hash to the same bucket. For each tagging action group $g_x \subseteq G$, we represent the categorical user attributes or item attributes as a boolean vector and concatenate it with $T_{rep}(g_x)$. We map $n$ vectors from a higher $(d + \sum_{i=1}^{|S_U|} \sum_{j=1}^{|a_i|} |a_i = v_j|)$ dimensional space for users (replace $|S_U|$ with $|S_I|$ for items) to a lower $d'$ dimensional space. Similar to Algorithm 1, we consider $l$ LSH hash functions and then post-process the buckets for satisfiability of the remaining constraints in order to retrieve the final result set of tagging action groups $G^{app}$ with maximum optimization score. Problem 1 in Table 1 has all three tagging action components set to similarity. In this case, we build one long vector for each tagging action group $g_x \subseteq G$ by concatenating boolean vector corresponding to categorical user attributes, boolean vector corresponding to categorical item attributes and numeric tag topic signature vector $T_{rep}(g_x)$. The dimensionality of the high-dimensional space for Problem 1 is $d + \sum_{i=1}^{S_U} \sum_{j=1}^{|a_i|} |a_i = v_j| + \sum_{i=1}^{S_I} \sum_{j=1}^{|a_i|} |a_i = v_j|$.

**Complexity Analysis:** The pre-processing time and search time of the complete LSH based algorithms continue to be $O(nld' \log n)$ and $O(n \log n)$ respectively. The space complexity of the algorithms is $O(nl)$.

Both SM-LSH-Fi and SM-LSH-Fo are efficient algorithms for solving TagDM similarity maximization problem instances and readily out-performs the baseline Exact, as shown in Section 6. However, there are other instantiations namely, Problems 4, 5 and 6 in Table 1 which concern tag diversity maximization. Since it is non-obvious how the hash function may be inversed to account for dissimilarity while preserving the properties of LSH, we develop another set of algorithms (less efficient than LSH based, as per complexity analysis) in Section 5 for diversity problems.

# 5. FDP BASED ALGORITHMS

The second of our algorithmic solutions borrows ideas from techniques employed in computational geometry, which model data objects as points in high dimensional space and determine a subset of points optimizing some objective function. Such geometric problem examples include clustering a set of points in euclidean space so as to minimize the maximum intercluster distance, computing the $k^{th}$ smallest or largest inter-point distance for a finite set of points in euclidean space, etc. Since we consider tagging action groups as tag signature vectors, and since the cardinality of the global set of topics (that, in turn, determines the size of each vector) is often high, computational geometry based approach is an intuitive choice to pursue.

We focus on a specific geometric problem, namely the facility dispersion problem (**FDP**), which is analogous to our TagDM problem instances, finding the tagging action groups maximizing the mining criterion. The facility dispersion problem deals with the location of facilities on a network in order to maximize distances between facilities, minimize transportation costs, avoid placing hazardous materials near housing, outperform competitors' facilities, etc. We consider the problem variant in Ravi et al.'s paper [18] that maximizes some function of the distances between facilities. The optimality criteria considered in the paper are MAX-MIN (i.e., maximize the minimum distance between a pair of facilities) and MAX-AVG (i.e., maximize the average distance between a pair of facilities). Under either criterion, the problem is known to be NP-hard by reduction from the Set Cover problem, even when the distances satisfy the triangle inequality [7]. The authors present an approximation algorithm for the MAX-AVG dispersion problem, that provides a performance guarantee of 4. The algorithm initializes a pair of nodes (i.e., facilities) which are joined by an edge of maximum weight and adds a node in each subsequent iteration which has the maximum distance to the nodes already selected.

The facility dispersion problem solution provides an approach to determine a set of tagging actions groups that have maximum average pair-wise distance, i.e., that are dissimilar in their tagging behavior. In fact, this approach may also be extended to determine a set of tagging action groups that are similar in their behavior, unlike the LSH based algorithm in Section 4 (which works only for similarity, not diversity). We consider each of the input $n$ tagging action groups as $d$-dimensional tag signature vector in a unit hypercube and intend to identify $k$ vectors with maximum average pairwise distance between them. We compare the input set $G$ of $n$ tagging action groups using a pairwise comparison function $F''_p(g_1, g_2, \texttt{tags}, diversity)$ that operates on tagging action group signature vectors; and return the set of tagging groups $\leq k$ having maximum diversity in tagging behavior.

Our FDP based algorithms work for Problems 4, 5 and 6 in Table 1 maximizing tag diversity. We first introduce an algorithm that returns the groups having maximum diversity in tagging behavior (Column $O$ in Table 1) and then discuss additional techniques to handle the multiple hard constraints in the solution (Column $C$ in Table 1).

## 5.1 Maximizing Diversity based on FDP

Our FDP based algorithm **DV-FDP** handles TagDM problem instances optimizing tag **DiV**ersity. Given an input set $G$ of $n$ tagging action groups, each having a numeric tag signature vector $T_{rep}(g_x)$, where $g_x \subseteq G$, we build the result set $G^{app}$ (we represent the result set as $G^{app}$ since the technique returns approximate solution) by adding a tagging action group in each iteration which has the maximum distance to the groups already included in the result set. Again, we use cosine similarity measure between two tag signature vectors for determining the distance since the distance metric hold triangular inequality property. Thus, our DV-FDP attempts to find one tight set of $k$ groups with maximum average pairwise distance between them. The approximation bounds for this algorithm follows from [18]:

THEOREM 4. *Let $I$ be an instance of the TagDM problem maximizing the mining criterion with $k \geq 2$ and no other hard constraints, where the collection of $n$ $d$-dimensional vectors are in a unit hypercube satisfying the triangle inequality. Let $G^{opt}$ and $G^{app}$ denote respectively the optimal set of $k$ tagging action groups returned by Exact and DV-FDP algorithms. Then $G^{opt}/G^{app} \leq 4$.*

Algorithm 2 is the pseudo-code of our DV-FDP algorithm. Once the $n \times n$ distance matrix $S^G$ is built using the cosine distance function, the implementation exhaustively scans $S$ for determining the best add operation in each of the subsequent iterations. If $A$ represents the result set, the objective is to find an entry from $G - A$ to add to $A$, such that its total sum of weight to a node in $A$ is maximum.

---

**Algorithm 2 DV-FDP ($G$, $O$, $k$):** $G^{app}$

*//Main Algorithm*
1: $S^G \leftarrow$ Compute $n \times n$ Distance Matrix(G)
2: $\{g_x, I_x, g_y, I_y\} \leftarrow \text{MAX}(S^G)$
3: $A \leftarrow [g_x, g_y]$
4: **while** $A \neq k$ **do**
5: $\quad g_z \leftarrow \Sigma_{\{z' \in [A], z \in [G-A]\}} \text{MAX}(S^{G-A})$
6: $\quad A \leftarrow [A, g_z]$
7: **end while**
8: $G^{app} \leftarrow A$
9: **return** $G^{app}$

---

**Complexity Analysis:** The complexity of the implementation of the DV-FDP algorithm is $O(n^2 + nk)$, i.e., $O(n^2)$ due to operations around the $n \times n$ distance matrix $S^G$. The space complexity of the algorithm is $O(n^2)$. Note that, our LSH based algorithms have better space and time complexity than FDP based algorithms. However, experiments in Section 6 show comparable execution time for LSH and FDP based algorithms in a practical setting.

Like SM-LSH, this algorithm does not leverage the hard constraints along user and item dimensions into the optimization solution, as well. We now illustrate approaches for including the multiple hard constraints into the solution.

## 5.2 Dealing with Constraints: Filtering

Similar to SM-LSH-Fi, a straightforward method of refining the result set of groups for satisfiability of all the hard constraints in TagDM problem instances is post-processing or **F**iltering. We refer to this algorithm as **DV-FDP-Fi**. Once the result set $G^{app}$ of $k$ groups is identified, we post-process it to retrieve the relevant answer set of tagging action groups, satisfying all the hard constraints. Note that, such post-processing of the result set for satisfiability of hard constraints may return null results frequently and hence we propose a smarter algorithm that folds some of the hard constraints into the DV-FDP approach, thereby decreasing the chances of hitting a null result.

## 5.3 Dealing with Constraints: Folding

In contrast to general DV-FDP algorithm whose objective is to add groups to the result set greedily so that average pair-wise distance is maximized, we want to retrieve the set in each iteration whose members, besides being dissimilar, satisfy many other constraints. In DV-FDP, the greedy add operation in Line 5 of Algorithm 2 maximizes tag diversity. If the algorithm includes a bad tagging action group to the result set in an iteration, the algorithm may return null result or an inferior approximate result, after final filtering of the result set for hard constraint satisfiability. Therefore, we propose our second approach in which multiple hard constraints are **F**olded into the add operation. We refer to this algorithm as **DV-FDP-Fo**. During each new group addition to the result set, we not only check for the pair with maximum distance, but also check for the satisfiability of the hard constraints $F'_p(g_1, g_2, \texttt{users}, m) \geq q$ and $F'_p(g_1, g_2, \texttt{items}, m) \geq r$, where $m \in \{\texttt{similarity}, \texttt{diversity}\}$. The algorithm terminates when the number of groups in result set equals $k$. Once the result set of $k$ groups is identified, we post-process the set for satisfiability of the support constraint, in order to retrieve the answer result of tagging action groups $G^{app'}$.

**Complexity Analysis:** The time and space complexity of the algorithm continues to be $O(n^2)$ in the worst case.

**Discussion:** Table 2 broadly summarizes our algorithmic contributions for solving the TagDM problem instances in Table 1. Note that, our algorithms are capable of handling *all* 112 concrete problem instances that our framework captures.

| Optimization | Algorithm | Constraints | Additional Techniques |
|---|---|---|---|
| similarity | LSH based | similarity | fold constraints |
| | | diversity | filter constraints |
| | | similarity, diversity | fold similarity constraints, filter diversity constraints |
| diversity | FDP based | similarity | fold constraints |
| | | diversity | fold constraints |
| | | similarity, diversity | fold constraints |

**Table 2: Summary of *TagDM* Problem Solutions.**

## 6. EXPERIMENTS

We conduct a set of comprehensive experiments for quantitative (Section 6.1) and qualitative (Section 6.2) analysis of our proposed algorithms for all 6 problems listed in Table 1. Our quantitative performance indicators are (a) *efficiency* of the algorithms, and (b) *analysis quality* of the results produced. The efficiency of our algorithms is measured by the overall response time, whereas the result quality is measured by the average pairwise distance between the $k$ tagging action group vectors returned by our algorithms (i.e., $F_{pa}$). In order to qualitatively assess the tagging behavior analysis generated by our approaches, we conduct a user study through Amazon Mechanical Turk as well as write interesting case studies.

**Data Set**: We use the MovieLens[3] 1M and 10M ratings dataset for our evaluation purposes. The MovieLens 1M dataset consists of 1 million ratings from 6000 users on 4000 movies while the 10M version has 10 million ratings and 100,000 tagging actions applied to 10,000 movies by 72,000 users. The titles of movies in MovieLens are matched with those in the IMDB dataset[4] to obtain movie attributes.

**User Attributes**: The 1M dataset has well-defined user attributes but no tagging information, whereas the 10M dataset has tagging information but no user attributes. Therefore, for each user in the 1M dataset with a complete set of attributes, we build her rating vector and compare it to the rating vectors (if available) of all 72,000 users in the 10M dataset. For every user in 10M dataset, we find the user in 1M dataset such that the cosine similarity of their movie rating vector is the highest (i.e., user rating behaviors are most identical). The attributes of user in 10M dataset are obtained from the closest user in 1M dataset. In this way, we build a dataset consisting of 33,322 tagging and rating actions applied to 6,258 movies by 2,320 users. The tag vocabulary size is 64,663. The user attributes are gender, age, occupation and zip-code. The attribute *gender* takes 2 distinct values: male or female. The attribute *age* is chosen from one of the eight age-ranges: under 18, 18-24, . . . , 56+. There are 21 different *occupations* listed by MovieLens such as student, artist, doctor, lawyer, etc. Finally, we convert zipcodes to states in the USA (or foreign, if not in USA) by using the USPS zip code lookup[5]. This produces the user attribute *location*, which takes 52 distinct values.

**Movie Attributes**: Movie attributes are genre, actor and director. There are 19 movie *genres* such as action, animation, comedy, drama, etc. The pool of actor values and director values, corresponding to movies which have been rated by at least one user in the MovieLens dataset, is huge. We pick only those actors and directors who belong to at least one movie that has received greater than 5 tagging actions. In our experiments, the number of distinct *actor* attribute values is 697 while that of distinct *director* is 210.

**Mining Functions**: The set of tagging action groups is built by performing a cartesian product of user attribute values with item attribute values. An example tagging action group is {gender=male, age=under 18, occupation=student, location=new york, genre=action, actor=tom hanks,

---

[3]http://www.grouplens.org/node/73
[4]http://www.imdb.com/interfaces
[5]http://zip4.usps.com

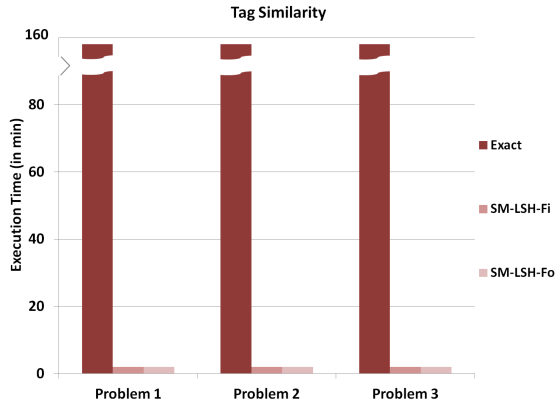
Figure 3: Execution Time:Problems 1, 2, 3 in Table 1

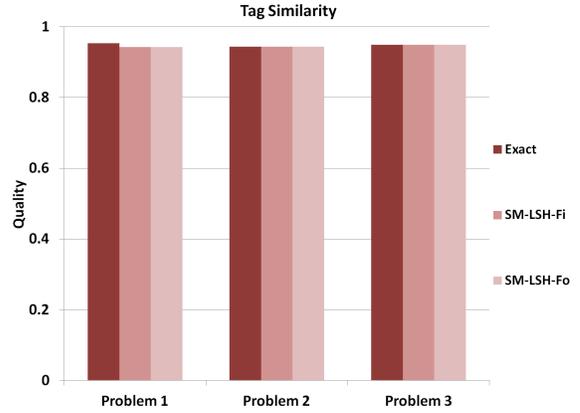
Figure 4: Quality:Problems 1, 2, 3 in Table 1

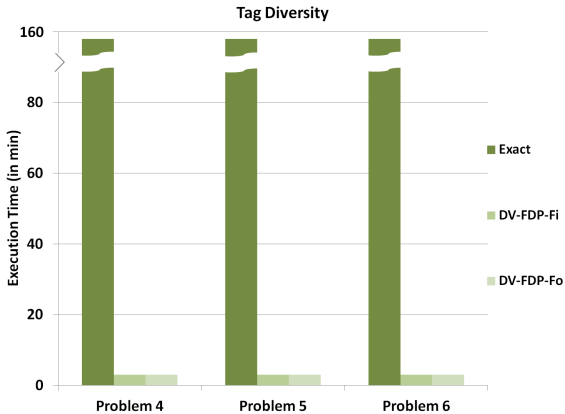
Figure 5: Execution Time:Problems 4, 5, 6 in Table 1

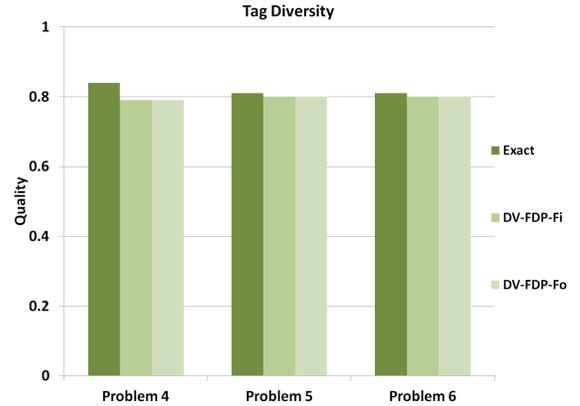
Figure 6: Quality:Problems 4, 5, 6 in Table 1

director=steven spielberg}. The total number of possible tagging action groups is more than 40 billion, while the number of tagging action groups containing at least one tuple is over 300K. For our experiments, we consider 4535 groups that contain at least 5 tagging action tuples. The user and item similarity (or diversity) is measured by determining the structural distance between user and item descriptions of groups respectively. For topic discovery, we apply LDA [3] as discussed in Section 2.1.2. We generate a set of 25 global topic categories for the entire dataset, i.e., $d = 25$. For each tagging action group, we perform LDA inference on its tag set to determine its topic distribution and then generate its tag signature vector of length 25. Finally, we use cosine similarity for computing pairwise similarity between tag signature vectors.

**System Configuration:** Our prototype system is implemented in Python. All experiments were conducted on an Ubuntu 11.10 machine with 4 GB RAM, AMD Phenom II N930 Quad-Core Processor.

### 6.1 Quantitative Evaluation

We compare the execution time of all 6 TagDM problem instantiations in Table 1 for the entire dataset (consisting of 33K tuples and 4K tagging action groups) using Exact, SM-LSH-Fi, SM-LSH-Fo, DV-FDP-Fi and DV-FDP-Fo algorithms. We use the name Exact for the brute-force approach on both tag similarity and diversity maximization instances. For all our experiments, we set the number of tagging action groups to be returned at $k = 3$, since the Exact algorithm is not scalable for larger k. Figure 3 and 4 compare the execution time and quality respectively of Exact and LSH based algorithms for Problems 1, 2 and 3 (Tag Similarity). Figure 5 and 6 compare the execution time and quality respectively of Exact and FDP based algorithms for Problems 4, 5 and 6 (Tag Diversity). The quality of the result set is measured by computing the average pair-wise cosine similarity between the tag signature vectors of the $k = 3$ tagging action groups returned. The group support is set at $p = 350$ (i.e., 1%); the user attribute similarity (or, diversity) constraint as well as the item attribute similarity (or, diversity) constraint is set to $q = 50\%$, $r = 50\%$ respectively. For LSH based algorithms, the number of hash tables is $l = 1$ while the initial value of $d'$ is 10.

We observe that the execution time of our algorithms is much faster than Exact, for both tag similarity and tag diversity problem instances. In Figure 3, the execution times of SM-LSH-Fi and SM-LSH-Fo for Problems 1, 2 and 3 are comparable to each other and is less that 1 minute. In Figure 5, the execution times of DV-FDP-Fi and DV-FDP-Fo for Problems 4, 5 and 6 are slightly more than 3 minutes. Despite significant reduction in execution time, our algorithms do not compromise much in terms of analysis quality, as evident from Figure 4 and Figure 6.

The number of input tagging action tuples available for tagging behavior analysis is dependent on the *query* under consideration. For the entire dataset, there are 33K such tu-ples. However, if we want to perform tagging behavior analysis of all movies tagged by {gender= male} or {location= CA}, the number of available tuples is 26,229 and 6,256 respectively. Or, if want to perform tagging behavior analysis of all users who have tagged movies having {genre= drama}, the number of tuples is 17,368. Needless to say, the number of tagging action tuples can have a significant impact on the performance of the algorithms since it affects the number of non-empty tagging action groups on which our algorithms operate. As a result, we build 4 bins having 30K, 20K, 10K and 5K tagging action tuples respectively (assume, each bin is a result of some query on the entire dataset) and compare our algorithm performances for one of the tag similarity maximization problems and one of the tag diversity maximization problems, say Problem 1 and Problem 6 from Table 1 respectively. Both Problems 1 and 6 have user and item dimension constraints set to similarity. Figures 7 and 8 compare the execution time and quality respectively of the Exact brute-force algorithm with our smart algorithms: SM-LSH-Fo for Problem 1 and DV-FDP-Fo for Problem 6. The group support is set at $p = 350$ (i.e., 1%); the user attribute similarity (or, diversity) constraint and the item attribute similarity (or, diversity) constraint are set to $q = 50\%$, $r = 50\%$ respectively, and $k = 3$. For each bin along the X axis, the first two vertical bars stand for Problem 1 (tag similarity) and the last two stand for Problem 6 (tag diversity).

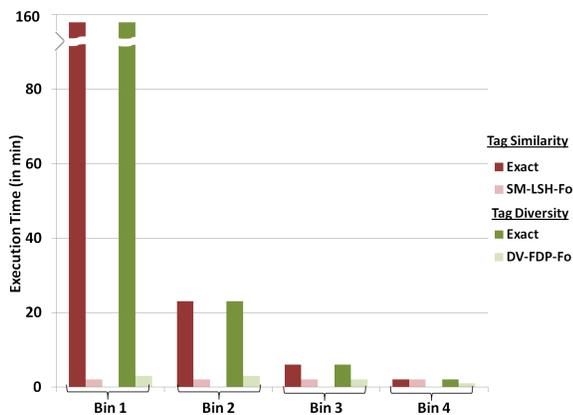

Figure 7: Execution Time:Varying Tagging Tuples

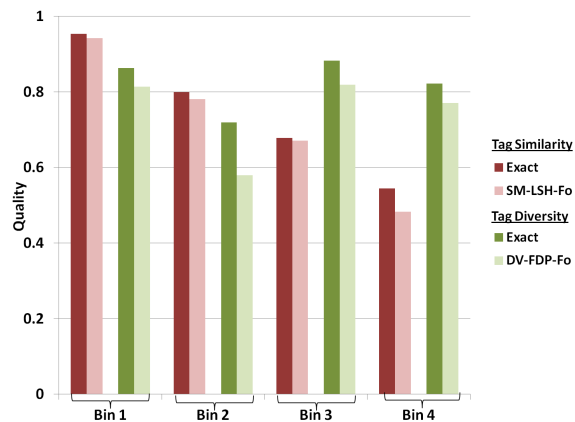

Figure 8: Quality:Varying Tagging Tuples

As expected, the difference in execution time between our algorithms and the Exact is small for bins with lesser number of tagging tuples for both tag similarity and diversity. However, our algorithms return results much faster than Exact for bins with larger number of tagging tuples. The quality scores continue to be comparable to the optimal answer, as shown in Figures 8.

## 6.2 Qualitative Evaluation

We now validate how social tagging behavior analysis can help users spot interesting patterns and draw conclusions about the desirability of an item, by presenting several anecdotal results on real data. We also compare the utility and popularity of the 6 novel mining problems in Table 1 in an extensive user study conducted on Amazon Mechanical Turk (AMT)[6].

[6] https://www.mturk.com

### 6.2.1 Case Study

We present few interesting anecdotal results returned by our algorithms for the following randomly selected queries:

1. *Analyze user tagging behavior for* {director= steven spielberg, genre= war} *movies:* Old male and young female use diverse set of tags for war movies "Saving Private Ryan" and "Schindler's List" directed by Steven Spielberg. This is because, the former is a movie about US military while the latter revolves around German military in World War II. Also, old male and young male tag "Schindler's List" dissimilarly: the former likes it while the latter does not.

2. *Analyze tagging behavior of* {gender= male, location= california} *users for movies:* Old male and young male living in California use similar tags for "Lord of the Rings" film trilogy of fantasy genre. However, they differ in their tagging towards "Star Wars" movies having similar genre. This is because, the genre of the latter series borders between fantasy and science fiction. Surprisingly, old male likes it while young male does not.

### 6.2.2 User Study

We conduct a user study through Amazon Mechanical Turk to elicit user responses towards the different TagDM problem instances we have focused on in the paper. We generate analysis corresponding to all 6 problem instantiations for the following randomly selected queries:

1. Analyze tagging behavior of {gender= male} users for movies.
2. Analyze tagging behavior of {occupation= student} users for movies.
3. Analyze user tagging behavior for {genre= drama} movies.

We have 30 independent single-user tasks. Each task is conducted in two phases: User Knowledge Phase and User Judgment Phase. During the first phase, we estimate the user's familiarity about movies in the task using a survey, besides her demographics. In the second phase, we ask users to select the most preferred analysis, out of the 6 presented to them, for each query. Responses from all users are aggregated to provide an overall comparison between all problem instances in Figure 9. The height of the vertical bars represent the percentage of users, preferring a problem instance. It is evident that users prefer TagDM Problems **2** (*find similar user sub-populations who agree most on their tagging*

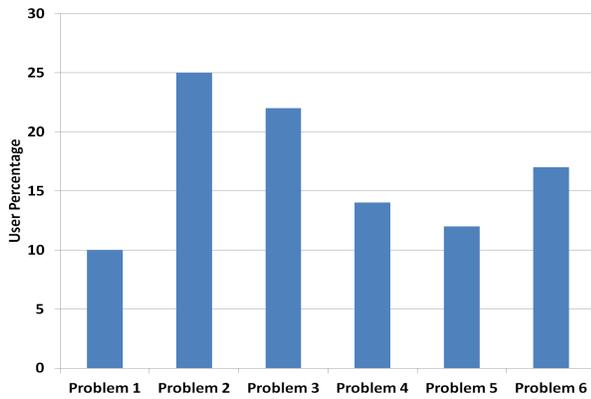

Figure 9: User Study

*behavior for a diverse set of items*), **3** (*find diverse user sub-populations who agree most on their tagging behavior for a similar set of items*) and **6** (*find similar user sub-populations who disagree most on their tagging behavior for a similar set of items*), having diversity as the measure for exactly one of the tagging component: item, user and tag respectively.

## 7. RELATED WORK

To the best of our knowledge, our work is the first to develop a general framework that encompasses mining collaborative tagging actions, studies its complexity and develops efficient algorithms. We summarize work related to topic discovery, tag mining and its applications, and the heuristics we use in our algorithms.

There are many topic discovery techniques such as tf*idf [19], Latent Dirichlet Allocation (LDA) [3, 2] and OpenCalais. In this work, we use LDA, a generative probabilistic method proven to be robust when looking for hidden topics in Web documents [3, 1].

Tag mining has been used in multiple applications including tag recommendations [17], item recommendations [16, 10], document navigation [11], and tagging motivation [15] However, most of these works are tailored to specific datasets and none of them defines a general mining problem, studies its complexity and develops efficient generic algorithms.

Locality Sensitive Hashing (LSH) and the Facility Dispersion Problem (FDP), were first introduced in [13, 8] and [12] respectively. LSH is used in prominent applications including duplicate detection and nearest neighbor queries [13]. In this work, we show how we adapt LSH to rank and choose the best bucket containing tagging analysis result. While being less efficient than LSH, the computational geometry based approach for the facility dispersion problem in [18] serves tag diversity problem instantiations and may be extended to solve similarity problems.

## 8. CONCLUSION

In this paper, we developed the first framework to mine social tagging behaviors. We identified a family of mining problems that apply two opposing measures: similarity and diversity, to the three main tagging components: users, items, and tags. We showed that any instance of those is NP-Complete and developed efficient algorithms based on locality sensitive hashing and solutions developed in computational geometry for the facility disperson problem. Our extensive experiments on the MovieLens dataset show the superiority of our algorithms over the brute-force approach. In the future, we plan to handle updates and insertions of new users, items and tags. We also intend to explore the applicability of our framework to other domains such as topic-centric exploration of tweets and news articles, an area that has been receiving a lot of attention lately. In particular, we would like to explore the usefulness of our techniques for mining and characterizing events in tweets and news.